\documentclass{iau}
\usepackage{graphicx}

\title[GR Modelling and AGN Jet-base]
{Pinpointing the base of the AGN jets through general relativistic
X-ray reverberation studies.}

\author[D. Emmanoulopoulos]
{D. Emmanoulopoulos$^1$}

\affiliation{
$^1$Physics and Astronomy, University of Southampton, \\ 
SO17 1BJ, Southampton, United Kingdom \\ 
email: {\tt D.Emmanoulopoulos@soton.ac.uk}
}

\pubyear{2014}
\volume{313} 
\pagerange{xx--xx}
\jname{Extragalactic jets from every angle}
\editors{F. Massaro, C.C. Cheung, E. Lopez, A. Siemiginowska, eds.}

\begin{document}

\maketitle

\begin{abstract}
Many theoretical models of Active Galactic Nuclei (AGN) predict that the X-ray corona, lying above the black hole, constitutes the base of the X-ray jet. Thus, by studying the exact geometry of the close black hole environment, we can pinpoint the launching site of the jet. Detection of negative X-ray reverberation time delays (i.e.\ soft band X-ray variations lagging behind the corresponding hard band X-ray variations) can yield significant information about the geometrical properties of the AGN, such as the location of the X-ray source, as well as the physical properties of the the black hole, such as its mass and spin. In the frame-work of the lamp-post geometry, I present the first systematic X-ray time-lag modelling results of an ensemble of 12 AGN, using a fully general relativistic (GR) ray tracing approach for the estimation of the systems' response functions. By combing these state-of-the art GR response models with statistically innovative fitting routines, I derive the geometrical layout of the 
close BH environment for each source, unveiling the position of the AGN jet-base.
\end{abstract}

\firstsection

\section{Introduction}
Based on the current paradigm, active galactic nuclei (AGN) contain a central black hole (BH) which is fed, in most cases, by an optically thick and geometrically thin accretion disc as a result of matter transportation inwards and angular momentum outwards. This disc radiates as a series of black-body components (\cite{shakura73}) with peak emission at optical-ultraviolet wavelengths (\cite{malkan83}). Part of this radiation is assumed to be Compton up-scattered within a mildly relativistic hot electron cloud, often called the \textit{X-ray corona}. This medium is often approximated as a point source (representing the centroid of the X-ray emitting source), lying above the central BH, on the axis of symmetry of the system (i.e.\ BH spin axis). This arrangement is known as the `\textit{lamp-post geometry}'.\par
The photons, from the X-ray corona, form a power-law spectrum, and depending on the location of the X-ray source, a substantial part of them may illuminate the accretion disc. In this case, the photons are either Compton scattered by free or bound electrons, or photoelectrically absorbed followed by fluorescent line emission, or by Auger de-excitation. This yields the so-called `reflection spectrum', consisting of a number of emission and absorption lines (mainly below 1 keV), together with the 6.4 keV Fe K$\alpha$ emission line from neutral material, which is the strongest X-ray spectral feature (\cite{george91}).\par
In this framework, detection of negative X-ray time delays, as a function of the Fourier frequency, between the soft band and the hard band photons, can shed light on the X-ray emission mechanism and the geometry of these systems. Under the lamp-post geometry assumption, in which the reflection of the hard X-rays occurs very close to the BH (few gravitational radii, $r_{\rm g}$), the negative X-ray time delays arise from the difference in the path-lengths between the soft (reflected photons) and the hard X-ray photons (coming directly from the X-ray source) to the observer.\par
The nature of the X-ray corona, however, is still very uncertain. Under the `accretion-disk jet connection' scenario (\cite{marscher02}), it could be the base of the jet below the point where the flow is highly relativistic (\cite{marscher06}). Currently the observed negative X-ray time-lags (i.e.\ soft-band variations lagging the hard-band variations; NXTL hereafter), has triggered a great deal of scientific interest on interpreting and modelling their nature. Modelling of NXTLs is usually done in terms of simple top-hat impulse response functions (THIRF) (\cite{miller10_1h0707,zoghbi11,emmanoulopoulos11b}). This approach is, however, just a parametrisation of the NXTL spectra and does not carry any physical information about the geometry (e.g.\ height of the X-ray source) and the physical properties of the BH (i.e.\ mass, spin).\par
In order to properly model the impulse response function, one has to take into account in detail the various general relativistic (GR) effects, affecting the geometric path 
of both the hard and reflected X-rays. In this paper, I describe briefly the results of \cite{emmanoulopouls14} (EMM14, hereafter) in which we perform the first systematic analysis of the time-lag spectra of a sample of 12 AGN using fully general relativistic impulse response functions (GRIRFs, hereafter). These functions are generated using a lamp-post model with variable BH mass, BH spin parameter, viewing angle and height of the X-ray source above the disc. Thus, assuming that the X-ray corona is the actual jet-base, these results can be used to locate its position with respect to the accretion disc.

\section{X-ray Observations}
For our analysis we have used 12 AGN observed by \textit{XMM}-\textit{Newton}. These AGN are: NGC\,4395, NGC\,4051, Mrk\,766, MCG--6-30-15, Ark\,564, 1H\,0707-495, IRAS\,13224-3809, ESO\,113-G010, NGC\,7469, Mrk\,335, NGC\,3516 and NGC\,5548. The \textit{XMM}-\textit{Newton} data were processed using {\sc scientific analysis system} ({\sc sas}) (\cite{gabriel04}) version 12.0.1. We consider only the EPIC pn data as they have a higher count rate and lower pile-up distortion than the MOS data. More information about the exact number of observations, for each source, and the data reduction procedures can be found in section~2 and Table~1 in EMM14.

\section{The Lamp-post Model}
The lamp-post model, that we consider in this paper, consists of the following three physical components: a central supermassive BH with an accretion disc illuminated by a point-like X-ray source located on the axis of the system. This system is characterised by the following parameters: spin parameter and mass of the central BH, height of the X-ray source, and viewing angle.\par
The accretion disc extends from the inner most stable circular orbit (ISCO), $r_{\rm in}$, to 1000 $r_{\rm g}$ and it is an optically thick, geometrically thin, Keplerian cold (i.e.\ neutral) disc. The X-ray source lies above the BH at height $h$ and it comprises the primary source of X-rays, which we assume to be static and to be emitting isotropically with a power-law spectrum of photon index $\Gamma=2$.\par 
The central BH is characterised by its spin parameter, $\alpha$, and its mass, $M$. For the former, which defines the ISCO, we consider three values: 0 (Schwarzschild BH, $r_{\rm in}=6$ $r_{\rm g}$), 0.676 (intermediate spin BH, $r_{\rm in}=3.5$ $r_{\rm g}$) and 1 (maximally rotating Kerr BH, $r_{\rm in}=1$ $r_{\rm g}$).\par
We have chosen a variety of heights for the X-ray source, depending on the spin of the BH. For the case of a Schwarzschild BH, we select an ensemble of 18 heights: $\{2.3,2.9,3.6,4.5,5.7,7,8.8,11,13.7,17.1,21.3,26.5,33.1,41.3,$ $51.5,64.3,80.2,100\}$ $r_{\rm g}$. For the intermediate case we add to the ensemble a lower height of 1.9 $r_{\rm g}$, and for the Kerr BH we add yet another height of 1.5 $r_{\rm g}$, respectively.\par
Finally, the system is observed by a distant observer at a viewing angle of $\theta$ i.e.\ $\theta=0$ or $90^{\circ}$ if the disc is face on or edge on, respectively. For each one of the three BH spin parameters and each one height we consider three angles: 20 40 and $60^{\circ}$.\par
This wide parameter space, consisting of the variables $\alpha$, $h$ and $\theta$, yields a total of $(20+19+18)\times 3=171$ different geometrical layouts of the lamp-post model.
The BH mass is not an additional variable in our estimation of GRIRFs as all time-scales and frequencies scale linearly with it, i.e.\ $t_{{\rm g},M}=4.9255 M\;\rm{s}$ where $M$ is given in units of $10^6$ M$_{\rm \odot}$. In each geometry the photons will follow different trajectories from the X-ray source to the disc and from the disc to the observer, and thus the response of the system will be different.

\section{The GRIRFs and the Time-lag Spectra and the Fitting procedure}
\subsection{Estimation of the GRIRFs}
In order to compute the response of the accretion disc to the primary illumination from the X-ray source (described by the power-law) we use a flare with a step function profile that has very short duration of 1 $t_{\rm g}$. The primary intrinsic spectrum has a normalisation of unity. Then we estimate the response of the disc by measuring the flux \textit{only} of the neutral fluorescent Fe K$\alpha$ line, at 6.4 keV in the rest frame of the accretion disc.\par
\begin{figure}
\includegraphics[width=2.645in]{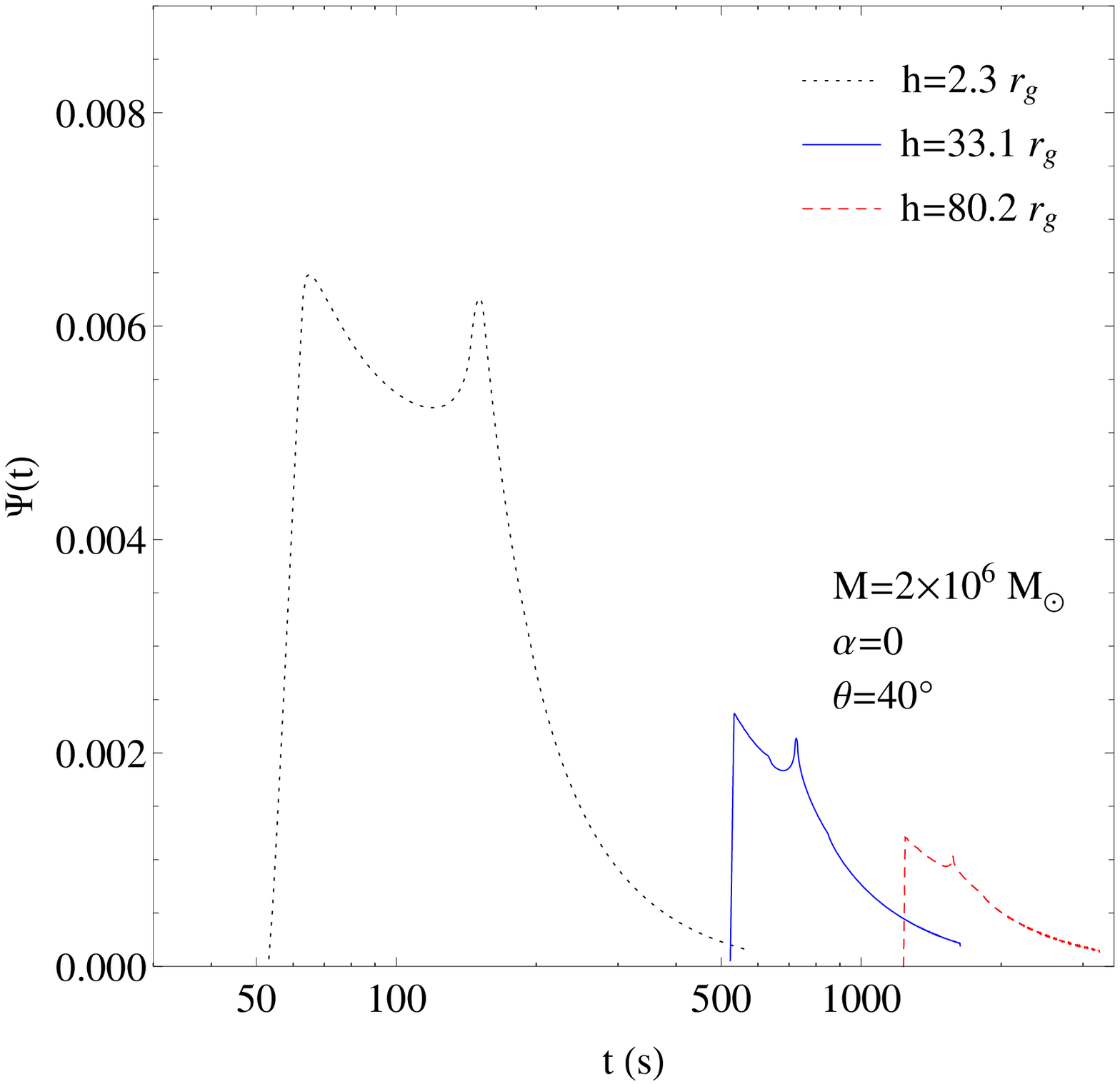} 
\includegraphics[width=2.645in]{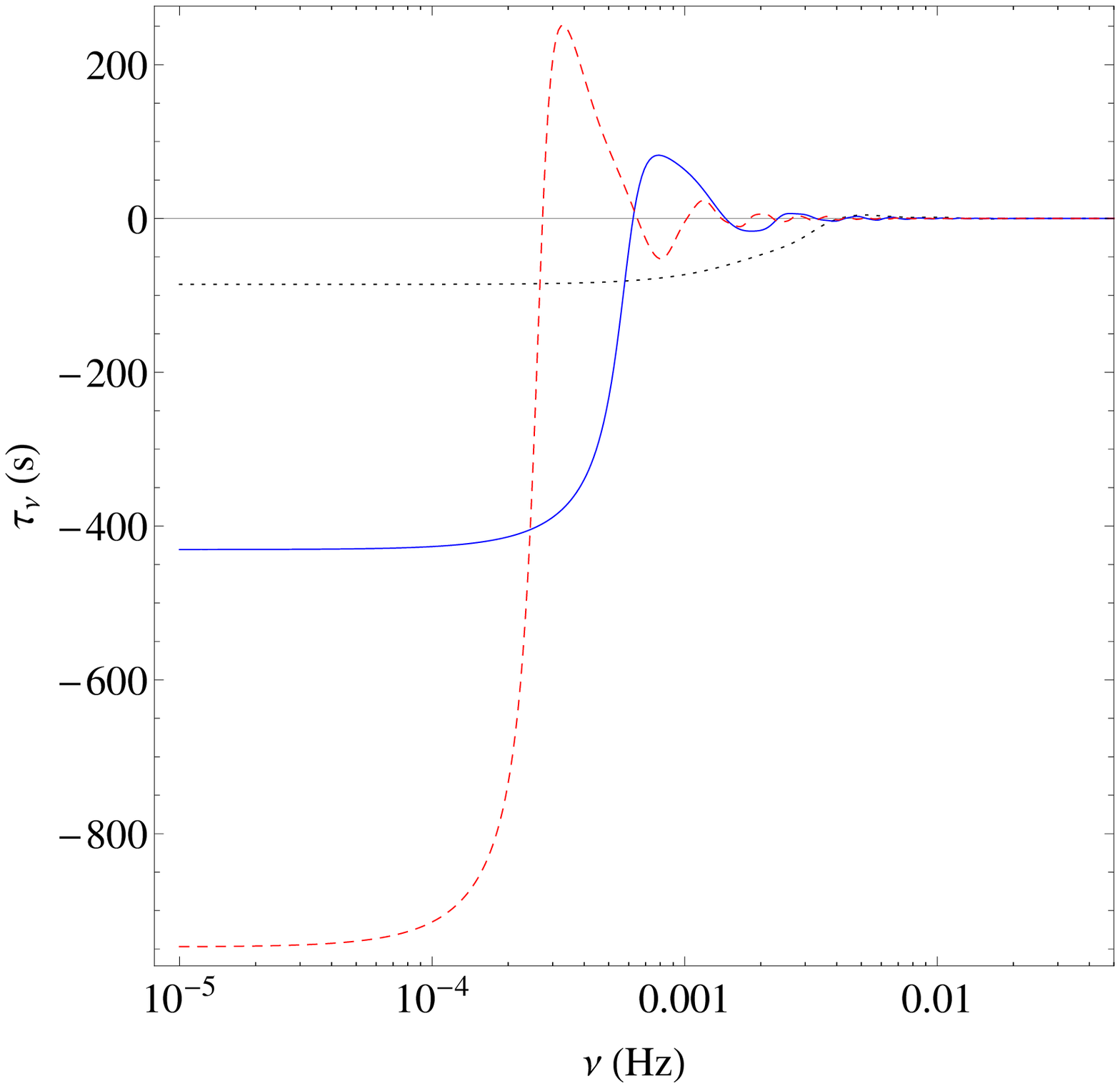}
\caption{[Left-hand panel] The GRIRF for the case of 3 different X-ray source heights. [Right-hand panel] The corresponding time-lag spectra as derived from the three GRIRF using  equation~\ref{eq:timeLagSpectra}}
\label{fig:grirfTlspec}
\end{figure}

To compute the Fe K$\alpha$ line flux we use the Monte Carlo multi-scattering code {\sc noar} (\cite{dumont00}), which takes into account both the direct and inverse Compton scattering processes. We assume a neutral accretion disc and an iron abundance equal to the Solar value. The resultant flux at the observer is computed using all the general relativistic effects (\cite{dovciak04}), without taking into account higher order images of the disc. Thus, we exclude photons emitted from the accretion disc (either from the top or the bottom surface) that go around the BH (any number of times), which reach the observer by travelling through the gap between the ISCO and the BH. Higher order images are more prominent for a Schwarzshild BH where the ISCO is the furthest out and the gap between the inner edge of the disc and the BH is the largest. Nevertheless, even in this case these photons do not contribute very much to the total flux. In the left-hand panel of Fig.~\ref{fig:grirfTlspec} we show the form of the 
GRIRFs 
for three different X-ray source heights.

\subsection{Theoretical time-lag spectra}
After estimating the GRIRFs, $\Psi(t)$, for each lamp-post model, we can construct the model time-lag spectra, $\tau_\nu$, at a given frequency $\nu$ following the procedure described in detail section 4.1 in EMM14.\par
Assume that the X-ray spectrum of AGN can be described by a power-law form $a(t)E^{-{\Gamma}}$, where $\Gamma$ is constant as a function of time and all the observed flux variability is due to the variations of the normalization $a(t)$ (primary continuum). Then, the source's hard X-ray emission in the 1.5--4 keV band, $h(t)$, which is dominated by the X-ray continuum source, can be written as 
\begin{eqnarray}
h(t)=b a(t)
\label{eq:hard_emis}
\end{eqnarray}
where $b=\int_{1.5\;{\rm keV}}^{4\;{\rm keV}}E^{-\Gamma}dE$. Let us then assume that $s(t)$ is the variable source's soft X-ray emission in the 0.3--1 keV energy range where, in addition to the continuum, we also detect emission from the reprocessing component so that
\begin{eqnarray}
s(t)=k a(t)+f\int_{0}^{\infty}\Psi(t^\prime)a(t-t^\prime)dt^\prime
\label{eq:soft_emis}
\end{eqnarray}
where $k=\int_{0.3\;{\rm keV}}^{1\;{\rm keV}}E^{-\Gamma}dE$ and $\Psi(t^\prime)$ 
After employing the cross-covariance function in our analysis (see EMM14), the time-lag spectrum, $\tau_\nu$, at a given frequency, $\nu$, is equal to 
\begin{eqnarray}
\tau_\nu=-\frac{\arg\left[1+(f/k)\int_{0}^{\infty}\Psi(t^\prime)e^{-i 2\pi\nu t^\prime} dt^\prime\right]}{2\pi\nu}
\label{eq:timeLagSpectra}
\end{eqnarray}
where $f$ describes potential differences between the actual observed flux of soft-band reflection spectrum, and that of the Fe K$\alpha$ line flux used for the modelling. In the right-hand panel of Fig.~\ref{fig:grirfTlspec} we show the form of the time-lag spectra as derived for the three GRIRFs corresponding to different heights.

\subsection{Observed time-lag spectra}
\label{ssect:observtlSpectra}
In order to estimate the time-lag spectra between two light curves we use we use the standard analysis method outlined in \cite{nowak99}. In brief, consider for a given source a soft and a hard light curve, $s(t)$ and $h(t)$, obtained simultaneously, consisting of the same number of $N$ equidistant observations with a sampling period $t_{\rm bin}$ (these are discretized and finite length versions of equations~\ref{eq:soft_emis} and \ref{eq:hard_emis}, respectively). For a given Fourier frequency, $f_j=j/(N t_{\rm bin})$ for $j=0,1,\ldots,[N/2-1\;\rm{or}\;(N-1)/2]$ (for even or odd $N$) we estimate the cross-spectrum, $\mathcal{C}(f_j)$ between the two light curves. Then, we average the complex cross-spectrum estimates, coming from all the observations, over a number of at least 10 consecutive frequency bins, yielding $m$ average cross-spectra estimates. Finally, for each average cross spectrum we derive its complex argument i.e.\ its angle with the positive real axis, known also as \textit{phase}, $\phi(f_{{\
rm bin},i})$ and we convert it to physical time units

\begin{eqnarray}
\tau(f_{{\rm bin},i})=\frac{\phi(f_{{\rm bin},i})}{2\pi f_{{\rm bin},i}}
\label{eq:tl_data}
\end{eqnarray}

Finally, from the cross-spectrum we estimate the coherence between $s(t)$ and $h(t)$ as a function of Fourier frequency. This quantity takes values between 0 and 1 and it is a measure of the linear correlation between the two light curves at a given Fourier frequency. In all our analysis we estimate the time-lag spectra down to $(3-5)\times10^{-3}$ Hz, but for the fitting procedure we consider only the time-lag estimates for which the coherence is greater that 0.15 corresponding to a physically meaningful phase correlation.

\begin{figure}
\includegraphics[width=2.645in]{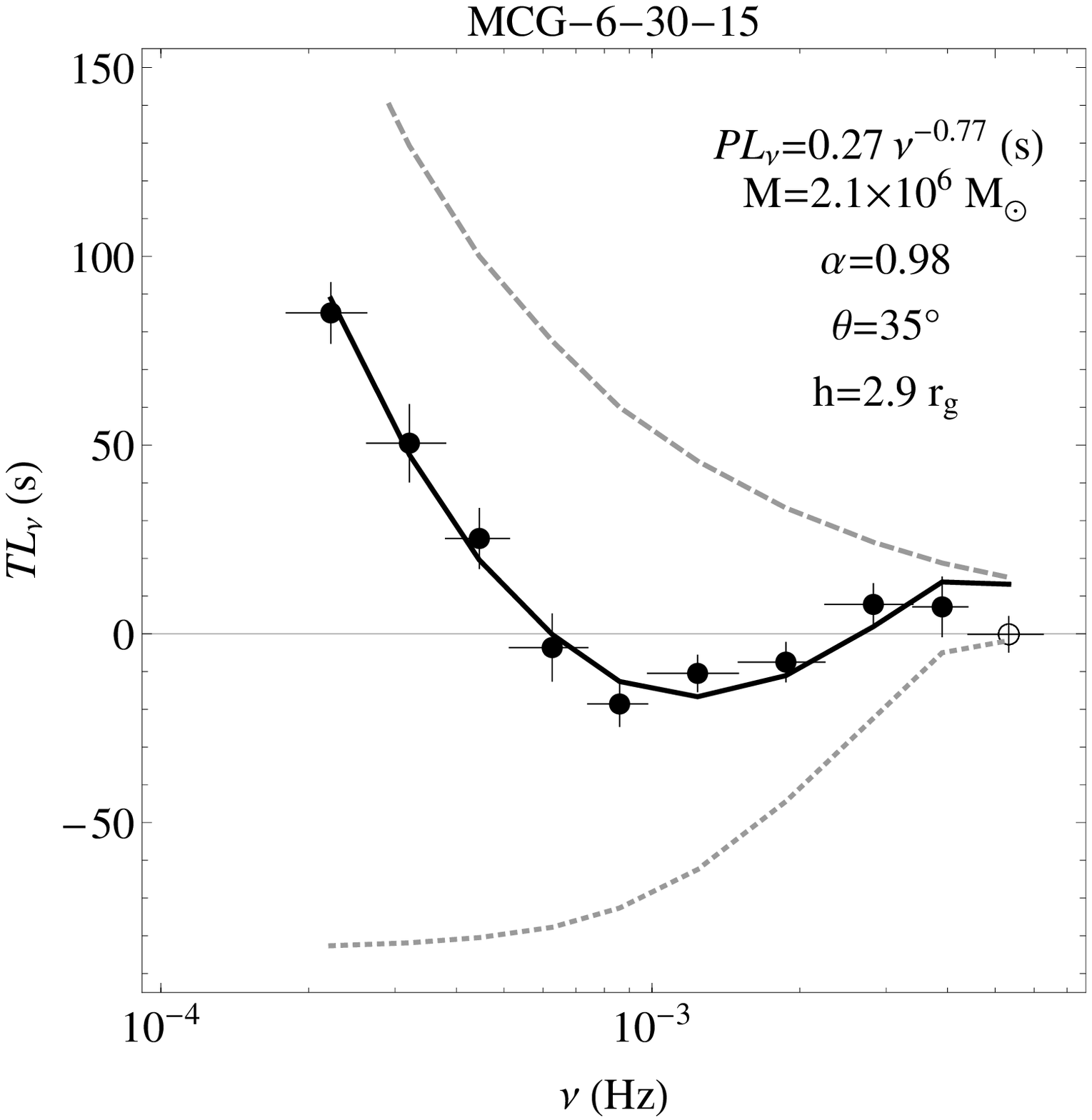} 
\includegraphics[width=2.645in]{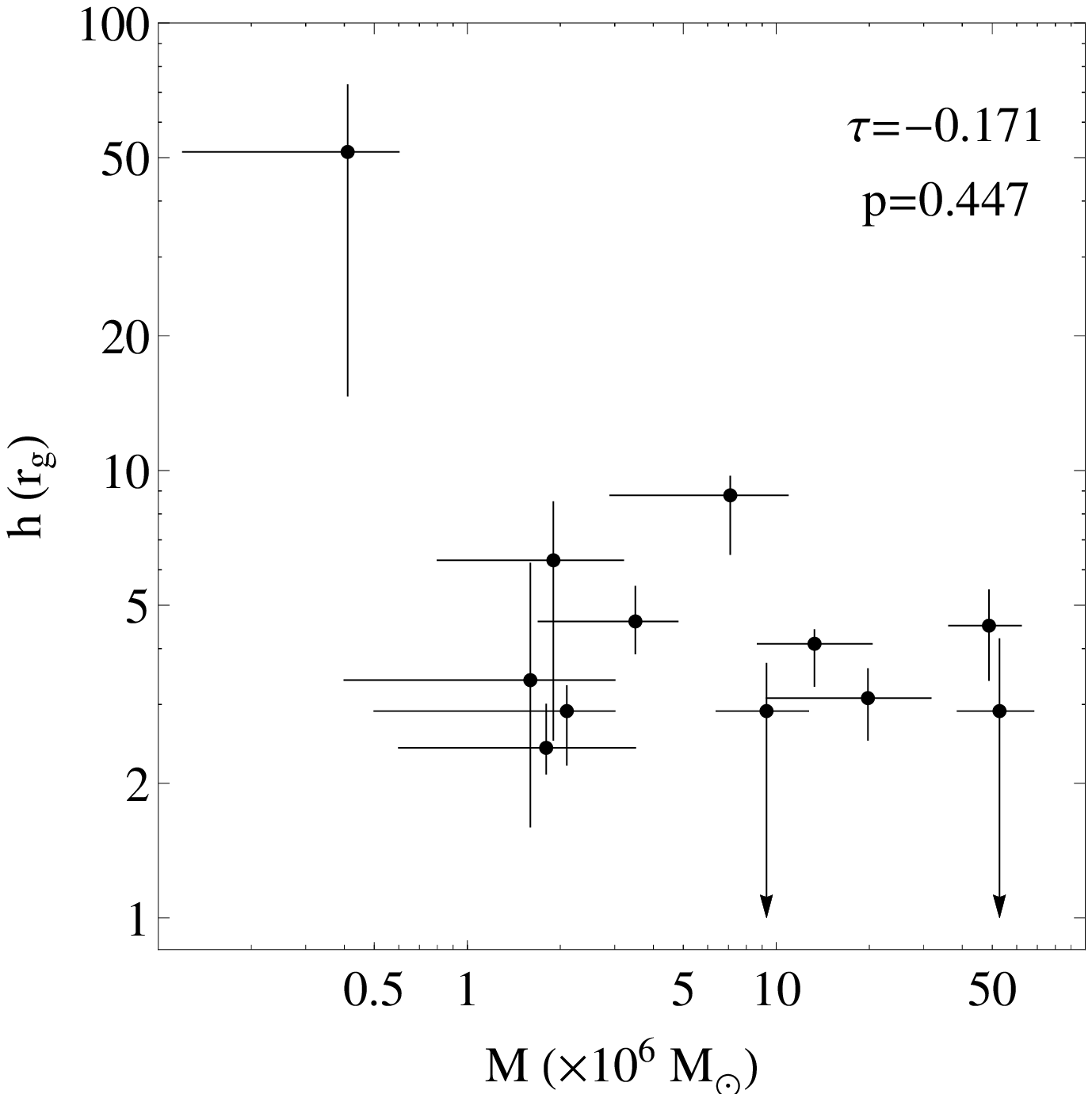}
\caption{[Left-hand panel] The best-fitting time-lag spectral model (black line) and the time-lag estimates for MCG--6-30-15, with coherence greater than 0.15 (filled circles), --open circles correspond to lower coherence values-- The two components of the best-fitting time-lag model are the GR reflected component (grey-dotted line) and the power-law (grey dashed line). [Right-hand panel] The best-fitting X-ray source heights versus the best-fitting BH mass for the 12 AGN. The parameter $\tau$ corresponds to the Kendall’s rank correlation coefficients and $p$ to the null-hypothesis, $H_0$, probability.}
\label{fig:mcg63015Heights}
\end{figure}

\subsection{The fitting procedure}
In order to fit the observed time-lag spectra, $\tau(f_{{\rm bin},i})$ binned into $m$ frequency bins (Section~\ref{ssect:observtlSpectra}), we require two time-lag spectral model components. The first one corresponds to the GR reflected component, estimated from equation~\ref{eq:timeLagSpectra} for a given GRIRF. This component, carries all the physical and geometrical information about the lamp-post model and gives rise predominantly to negative lags, particularly at the low frequency range. During the fit, we fix $f$ to 0.3, so that the total emission flux of the Fe K$\alpha$ line, over all energies, is 30 per cent that of the input X-ray continuum spectral flux. The second component consists of a simple power-law, $PL_\nu(A,s)=A\nu^{-s}$, providing us with positive time-lags. During the fitting procedure, we discretize and average the overall model in the same way as we do with the observed time-lag spectra and we use a grid in the five-dimensional parameter space, $\mathbf{v}=\{M,\alpha,\theta,h,A,s\}$. 
Then, for each averaged discretized overall time-lag spectral model, within each grid-cell, $k$, we estimate the squared differences between the model and the data using a $\chi^2$ estimator as the merit function. Details for the fitting procedure i.e.\ grid, discretization, and minimization can be found in section~5 of EMM14. In the left-hand panel of Fig.~\ref{fig:mcg63015Heights} we show the best-fitting model together with the best-fitting parameter values for the case of MCG--6-30-15. The same procedure is repeated for the ensemble of the 12 AGN (fig.~8 in EMM14).

\section{Results and Discussion}
Through our modelling, we are able to derive the best-fitting heights of the X-ray corona for our AGN ensemble consisting of 12 sources (Fig.~\ref{fig:mcg63015Heights}, right-hand panel)\footnote{Through our modelling we also derive the BH mass/spin, the viewing angle of the sources and the parameters of the power-law component, i.e.\ normalization and index (see EMM14).}. The average X-ray source height is $3.72^{+0.56}_{-0.52}$ $r_{\rm g}$, including the first estimate of NGC\,4395 (first point) coming from a very poor fit (reduced $\chi^2$ of 6.24). Assuming that the X-ray source corresponds to the jet-base, then AGN-jets should be launched very close to the BH.\par
Apart from the `accretion-disk jet connection' scenario, which is supported by radio, optical and X-ray data (\cite{chatterjee11}), AGN show signatures of jet structure similar to the ones observed in blazars purely from X-ray data, e.g.\ the `harder when brighter' behaviour of NGC\,7213 (\cite{emmanoulopoulos12}). The lamp-post model that we consider in our analysis, depicts the bulk of the X-ray emission (i.e.\ centroid) and not an actual X-ray emitting point above the accretion disc. More complex geometries (e.g. discs above or on top of the accretion disc, outflows) can be considered, but based on the quality of the current data and the number of time-lag estimates, it is impossible to distinguish differences in different geometrical scenarios.

\begin{acknowledgements}
DE acknowledges the Science and Technology Facilities Council (STFC) for support under grant ST/G003084/1. This research has made use of NASA's Astrophysics Data System Bibliographic Services. Finally, I would like to thank the conference organisers for covering my travel expenses.  
\end{acknowledgements}

\end{document}